\documentclass[aps,twocolumn,showpacs,superscriptaddress,floatfix,letter]{revtex4}
\usepackage{graphicx}
\usepackage{psfrag}
\usepackage{color}
\begin{document}
\newcommand{\beq}{\begin{equation}}
\newcommand{\eeq}{\end{equation}}
\newcommand{\ben}{\begin{eqnarray}}
\newcommand{\een}{\end{eqnarray}}
\newcommand{\baq}{\begin{array}}
\newcommand{\eaq}{\end{array}}
\newcommand{\om}{(\omega )}
\newcommand{\bef}{\begin{figure}}
\newcommand{\eef}{\end{figure}}
\newcommand{\leg}[1]{\caption{\protect\rm{\protect\footnotesize{#1}}}}
\newcommand{\ew}[1]{\langle{#1}\rangle}
\newcommand{\be}[1]{\mid\!{#1}\!\mid}
\newcommand{\no}{\nonumber}
\newcommand{\etal}{{\em et~al }}
\newcommand{\geff}{g_{\mbox{\it{\scriptsize{eff}}}}}
\newcommand{\da}[1]{{#1}^\dagger}
\newcommand{\cf}{{\it cf.\/}\ }
\newcommand{\ie}{{\it i.e.\/}\ }
\setlength\abovedisplayskip{5pt}
\setlength\belowdisplayskip{5pt}

\title{Interplay of superradiance and disorder in the Anderson Model}

\author{G.~L.~Celardo}
\affiliation{Dipartimento di Matematica e
Fisica and Interdisciplinary Laboratories for Advanced Materials Physics,
 Universit\`a Cattolica, via Musei 41, 25121 Brescia, Italy
and\\
 Istituto Nazionale di Fisica Nucleare,  Sezione di Pavia, 
via Bassi 6, I-27100,  Pavia, Italy}
\author{A.~Biella}
\affiliation{Dipartimento di Matematica e
Fisica,
 Universit\`a Cattolica, via Musei 41, 25121 Brescia, Italy}
\author{L.~Kaplan}
\affiliation{Tulane University, Department of Physics, New Orleans, Louisiana 70118, USA}
\author{F.~Borgonovi}
\affiliation{Dipartimento di Matematica e
Fisica and Interdisciplinary Laboratories for Advanced Materials Physics,
 Universit\`a Cattolica, via Musei 41, 25121 Brescia, Italy
and\\
 Istituto Nazionale di Fisica Nucleare,  Sezione di Pavia, 
via Bassi 6, I-27100,  Pavia, Italy}
\begin{abstract}                

Using a non-Hermitian Hamiltonian
approach to open systems, we study the interplay of disorder and 
superradiance in a one-dimensional Anderson model.                         
Analyzing the complex eigenvalues of the non-Hermitian Hamiltonian,
a transition to a superradiant regime is shown to occur.
As an effect of openness the structure of 
 eigenstates  undergoes a strong change in the                       
superradiant regime: we show that the sensitivity to disorder
of the superradiant and the subradiant subspaces is very different;
superradiant states remain delocalized as disorder
increases,  while subradiant states are sensitive to the degree of disorder. 
\end{abstract}                                                               
                                                                            
\date{\today}
\maketitle

\maketitle

\section{Introduction}
Nanoscopic systems in the quantum coherent regime
are at the center of many research fields
in physics, ranging from quantum computing and cold atoms to transport in nanoscale and mesoscopic systems.
Transport in the quantum coherent regime
can be considered one of the central subjects in
modern solid state physics~\cite{Beenakker,Lee} 
and in cold atom physics~\cite{kaiser}.
Transport properties depend strongly on the degree of openness of the system.

As a consequence of quantum coherence many interesting features arise.
Here we focus on two important effects induced by quantum coherence:
Anderson localization~\cite{Anderson} and Dicke superradiance~\cite{dicke54}.
Anderson Localization is driven by intrinsic disorder and 
consists in a suppression
of diffusion due to an exponential localization  of 
the eigenfunctions of the system.
Dicke superradiance is driven by the
fact that the system is open, namely  coupled to an external
environment characterized by a continuum of states.
To explain the superradiance effect, consider a discrete quantum system
 coupled to an environment having a continuum of states.
The system-environment coupling alters the unperturbed energy levels: 
it causes an energy shift and the appearance of a resonance
width (inverse lifetime) for each level. For weak coupling strength,
all resonance widths  are roughly equal. However,
once the coupling strength reaches a critical value, the 
widths start to overlap, and width segregation occurs. 
In this regime, almost the entire (summed up) decay width is
 allocated to just a few short-lived ``superradiant'' states, 
while all other states are long-lived 
(and effectively decoupled from the environment). 
We call this segregation the  ``superradiance transition'' (ST).

In the superradiant regime, the effect of the opening is large,
and cannot be treated perturbatively.
Thus, a consistent way to take
the effect of the opening into account for arbitrary coupling strength
between the system and the outside world is highly desirable.
The effective non-Hermitian Hamiltonian approach to open quantum
systems has been shown to be a very effective tool in addressing
this issue~\cite{heff}.
Non-Hermitian Hamiltonians have been already employed to study
realistic open quantum systems, such as transport through
quantum dots~\cite{qdots}, superradiance in cold atoms~\cite{kaiser} and
nuclear physics~\cite{zelenuclear}.
The superradiance effect
has also been studied using random matrix theory
\cite{verbaarschot85,puebla},
in microwave billiards~\cite{rotter2} and in paradigmatic models
of coherent quantum transport~\cite{kaplan,rottertb}.
As an example of the importance
of the ST, maximum transmission in a realistic model for quantum transport 
was shown to be achieved exactly at the ST~\cite{kaplan}.

In this paper we analyze a one-dimensional Anderson model, where a particle
hops from site to site in the presence of disorder, 
and is also allowed to escape the system from any site.
When the wavelength of the particle is comparable with the sample size,
an effective long-range hopping is created between the sites. 
This coupling can induce the ST, which affects in 
a non-trivial way the transport properties of the system.
Similar models of quantum transport with coherent
 dissipation have been already considered in the literature~\cite{jung1}, but
a detailed analysis of the interplay of
 localization and superradiance has been lacking.
This interplay has been recently analyzed
 in Ref.~\cite{kaplan}, but there the particle was allowed to escape only
from the end sites, while in the situation
 analyzed in this work, all sites are coupled to
the external environment.
This situation occurs in many important physical situations, such as
 in cold atoms, where a single photon is injected in the atomic cloud~\cite{kaiser}, or in quantum dots~\cite{exp}. 
 
Intrinsic  disorder and opening to the environment have opposing effects:
while disorder tends to localize
 the wave functions, the opening tends to delocalize
them, since it induces a long range interaction.
The aim of this paper is to study the interplay of disorder and 
opening, and the relation to superradiance.
We show that while below the ST, all states
 are affected by the disorder and the opening in a similar way,
above it, the effects are quite different for superradiant
 and subradiant subspaces, the latter being more affected
by disorder than the former.

In Sec.~\ref{sec-model} we introduce the model, in Sec.~\ref{sec-st} we analyze the 
ST in our system, and
in Sec.~\ref{sec-results} we present our main numerical results, which we partly
 justify in Sec.~\ref{sec-pt} using perturbation theory.
Finally in Sec.~\ref{sec-conc} we present our conclusions.

\section{Model}
\label{sec-model}

Our starting point is the standard one-dimensional Anderson 
model~\cite{Lee,Anderson}, for the motion of a particle
in a disordered potential.
The Hamiltonian of the Anderson model can be written as: 
\begin{equation}
H_0= \sum_{j=1}^{N} E_j | j\rangle \langle j| + \Omega \sum_{j=1}^{N-1} \left(| j \rangle \langle j+1|
+| j+1 \rangle \langle j|\right) \,,
\label{AM}
\end{equation}
where $E_j$ are random variables uniformly distributed
in $[-W/2 ,+W/2]$,  $W$ is a disorder parameter,
 and $\Omega$ is the tunneling transition amplitude
(in our numerical simulations we set $\Omega=1$).
For $W=0$ the eigenstates are extended and we have for the eigenvalues: 
\beq
E_{q} = -2\Omega \cos\left( \frac{\pi q}{N+1} \right), 
\eeq
and the eigenstates:
\beq
\psi_q(j)= \sqrt{\frac{2}{N+1}} \sin \left( \frac{\pi q}{N+1}j \right) \,,
\eeq
where $q=1, ...,N$ is a quantum number and $j=1, ...,N$ is a 
discrete coordinate.
In this case, the eigenvalues
lie in the interval $[-2\Omega, 2\Omega]$,
so the mean level spacing can be estimated as $D=4\Omega/N$.
For $W \ne0$, the eigenstates of the one-dimensional Anderson model
are exponentially localized on the system sites, with exponential
tails given by $|\psi(j)| \sim \exp(-|j-j_0|/\xi)$, where for weak disorder,
the localization length $\xi$ can be written as:
\begin{equation}
\xi \approx 96 \ (1-(E/2\Omega)^2)\left(\frac{\Omega}{W }\right)^{2} \,.
\label{loc}
\end{equation}
For $E=0$, Eq.~(\ref{loc}) has to be modified 
and we have~\cite{Felix}: 
$$\xi \approx 105.2 \ \left(\frac{\Omega}{W}\right)^{2} \,.$$

The phenomenon of Anderson localization
was studied in a closed disordered chain, while
in our case we can vary the degree of openness of the system.
In particular we consider the model 
in which all the sites are coupled to a common channel
in the continuum, with equal coupling strength $\gamma$.
This situation can arise when the wavelength of the decaying particle
is much larger than the size of the system. 
This results in a coherent dissipation, which
differs from the usual dissipation where every site decays independently
to a different channel in the continuum. A comparison
between these two different mechanisms will be the subject of a 
future work. 
The continuum coupling can be taken into account with the aid of an 
effective non-Hermitian Hamiltonian~\cite{kaplan},
which in general can be written 
as,  $$H_{\mathrm{eff}} (E) = H_0 + \Delta (E) -i Q(E)\,,$$ 
where $H_0$ is the Hermitian Hamiltonian of the closed system
decoupled from the environment and 
$\Delta (E)$ and $Q(E)$ are the induced energy shift 
and the dissipation, respectively.
Neglecting the energy dependence and the energy shift
we have 
\begin{equation}
(H_{\mathrm{eff}})_{ij}=(H_0)_{ij} -\frac{i}{2}  \sum_c A_i^c (A_j^c)^* \,,
\label{Heff}
\end{equation} 
where $A_i^c$ are the transition amplitudes
from the discrete states $i$ to the continuum channels $c$.

In the case under study, we have only one decay channel, $c=1$, and all couplings
are equal, so that
$A_i^1= \sqrt{\gamma}$. Thus
 the effective Hamiltonian can be written as:
\begin{equation}
H_{\rm eff}= H_0 -i\frac\gamma2 Q \,,
\label{amef}
\end{equation}
where $H_0$ is the
 Anderson Hamiltonian with diagonal disorder, Eq.~(\ref{AM}),
 and $Q_{ij}=1$ $\forall i,j$.

In order to study the interplay of Anderson 
localization and superradiance we
analyze the participation ratio ($PR$) of the eigenstates 
of $H_{\rm eff}$, defined as,
 \beq
 PR= \left\langle{\frac{1}{ \sum_i |\langle i| \psi \rangle|^4}}\right\rangle \,,
 \label{pr}
 \eeq
 where the average is over disorder.
 For completely extended states we have $PR=N$ and for completely localized states we have $PR=1$.


\section{Superradiance transition}
\label{sec-st}
\begin{figure}[htbp]
\centering
\includegraphics[scale=0.4]{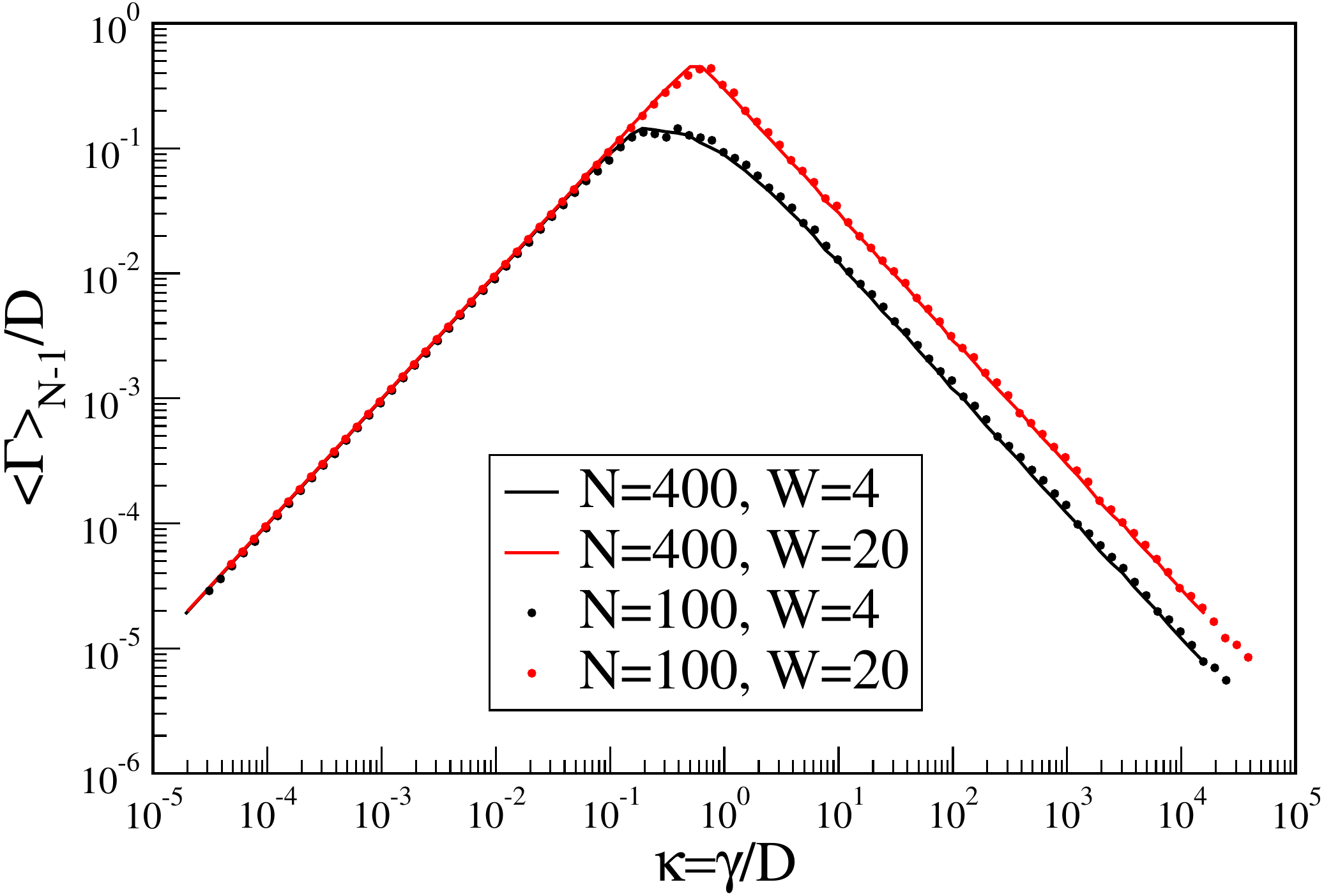}
\caption{The average width of the $N-1$ subradiant states, normalized by the mean level spacing $D$, 
versus the effective coupling 
strength $\kappa$ for different values of $N$ and $W$, and $\Omega=1$.
Here we average over $100$ disordered configurations.}
\label{ave2}
\end{figure}

ST can be analyzed by studying 
the complex eigenvalues 
${\cal E}_r = E_r -i \Gamma_r /2$ of $H_{\rm eff}$
defined in Eq.~(\ref{amef}). 
As the coupling between the  states and the continuum increases, 
one observes a rearrangement of the widths $\Gamma_r$.
ST is expected to occur for $\langle \Gamma \rangle /D \simeq 1$~\cite{heff,kaplan}.  
The average width $\langle \Gamma \rangle$ is $\gamma$, so
we can define
\beq
\kappa=\gamma/D
\eeq
as the parameter controlling the 
coupling strength to the continuum. 
In the deep localized regime where disorder is strong ($W \gg \Omega$) we can write $D \approx W/N$, so that the effective coupling strength can be written as:
\beq
\kappa=\frac{ \gamma N}{W}
\label{k}
\eeq
In Fig.~\ref{ave2} we show that ST
occurs at $\kappa \sim  1$ 
for different values of $W/\Omega$ and $N$.

For $\kappa \gg 1$, we can treat the matrix $Q$ as the leading term in Eq.~(\ref{amef}), and $H_0$ as a perturbation.
The superradiant state $|SR\rangle$ is given to zeroth order by the only
eigenstate of $Q$ with nonzero eigenvalue:
 $|d\rangle =\frac1{\sqrt{N}}(1,...,1)^T$, and the energy of $|SR\rangle$
is evaluated at first order as
\begin{equation}
\langle d| H_{\rm eff}|d \rangle =\epsilon -i \frac\gamma2N \,,
\end{equation}
where 
$$\epsilon=\frac1N \sum_{i=1}^N E_i+ 2\Omega \frac{N-1}N$$
 and $E_i$ are the random diagonal elements of $H_0$.
Averaging over disorder and
taking into account that $E_i$ are distributed uniformly
in
  $[-W/2,W/2]$ we obtain,
\begin{equation}
\label{meanen}
\langle \epsilon \rangle=2 \Omega \frac{N-1}N 
\end{equation}
and
\beq
\label{variance2}
{\rm Var}(\epsilon) = \langle \epsilon^2\rangle - \langle{\epsilon}\rangle^2 = \frac{W^2}{12N} \,.
\eeq
These results agree with our numerical simulations
 for different values of $N$ and 
allow one to know the position in the energy  band
of the superradiant state in the limit $\kappa\gg1$.
From Eq.~({\ref{meanen}}) we deduce that the mean energy 
$\langle{\epsilon}\rangle$ of the superradiant state is independent of $W$.

\section{Numerical Results}
\label{sec-results}
\begin{figure}[htbp]
\centering
\includegraphics[scale=0.4]{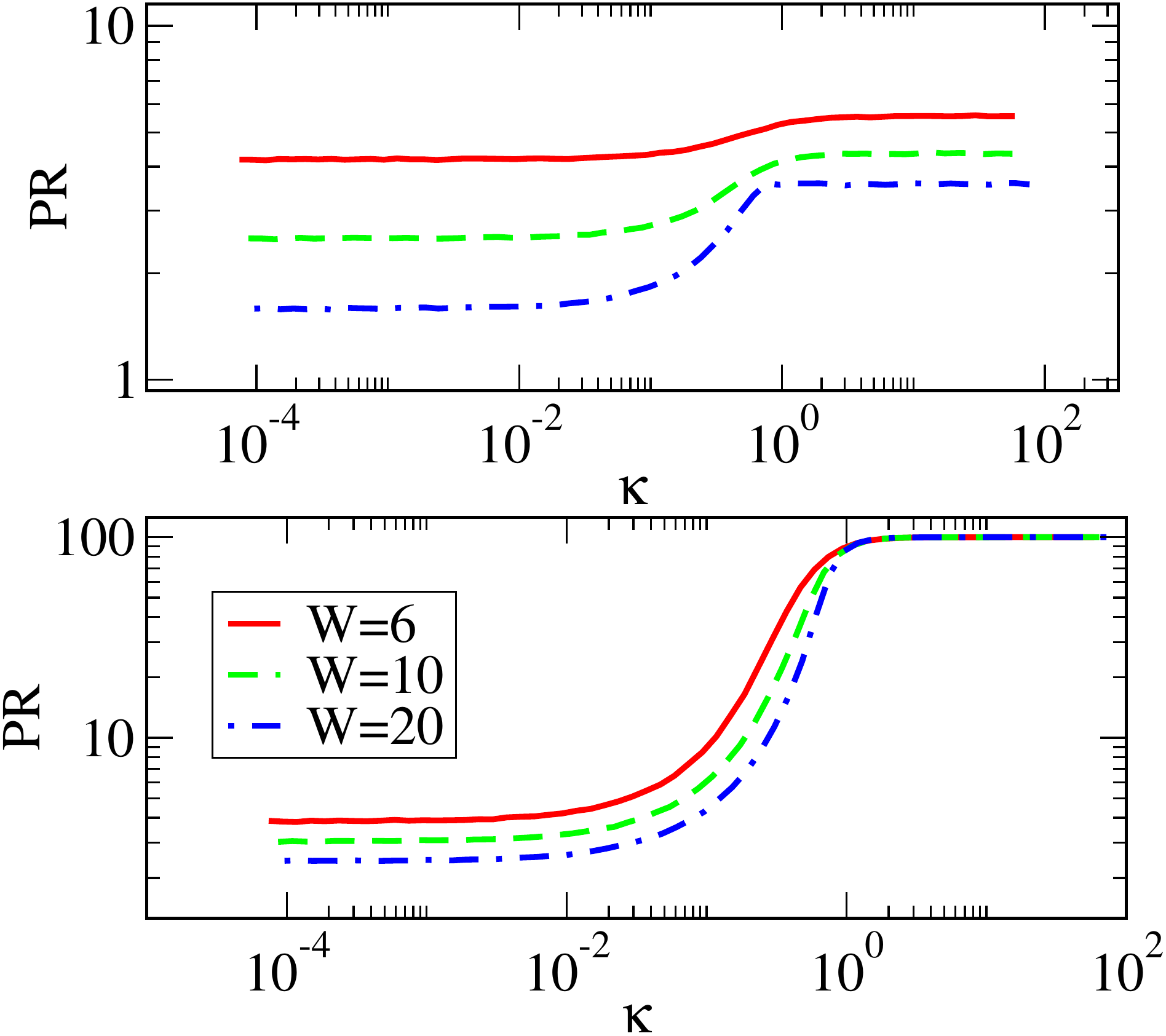}
\caption{The participation ratio $PR$ is shown as a function of $\kappa$ for different disorder strengths. In the upper panel  we consider states with $-1.5 \le E/\Omega \le -0.5$, which
become subradiant for large $\kappa$, while in the lower panel we consider the state with the largest width, which corresponds to the superradiant
state for large $\kappa$. Here $N=100$, $\Omega=1$, and the $PR$ is averaged over $4000$ 
disorder realizations.  }
\label{PRsubsr}
\end{figure}

In order to study the interplay of superradiance and disorder we
 have analyzed the $PR$ of the eigenstates
 of the non-Hermitian Hamiltonian, Eq.~(\ref{amef}). 
As explained in the previous section, 
as the coupling with the continuum is increased
we have the formation of one superradiant state (the one with the
largest width) and $N-1$ subradiant ones.
In Fig.~\ref{PRsubsr} (upper panel) we analyze the $PR$ as a function of $\kappa$ for 
the states that become subradiant for $\kappa > 1$, and in 
 Fig.~\ref{PRsubsr} (lower panel)
we analyze the case of  
the state with the largest width, which becomes superradiant 
for $\kappa > 1$.
As the opening, determined by the parameter 
$\kappa$, increases, the $PR$ of both superradiant and subradiant states
increases, showing that the opening has a delocalizing effect. 
But the consequences of the opening  
are very different for superradiant and subradiant states. 
For the latter, the $PR$ reaches a plateau value above 
the ST ($\kappa \approx 1$), which is 
slightly higher than the $PR$ for $\kappa \ll 1$.  Moreover 
on increasing the  disorder, the $PR$ of the subradiant states
 decreases, both below and above the ST, see Fig.~\ref{PRsubsr} upper panel. 
The situation is different for the superradiant states. 
Above the ST these states become completely delocalized
($PR=N$) and their delocalization is not affected by an increase in $W$,
see Fig.~\ref{PRsubsr} lower panel.  

\begin{figure}[htbp]
\centering
\includegraphics[scale=0.4]{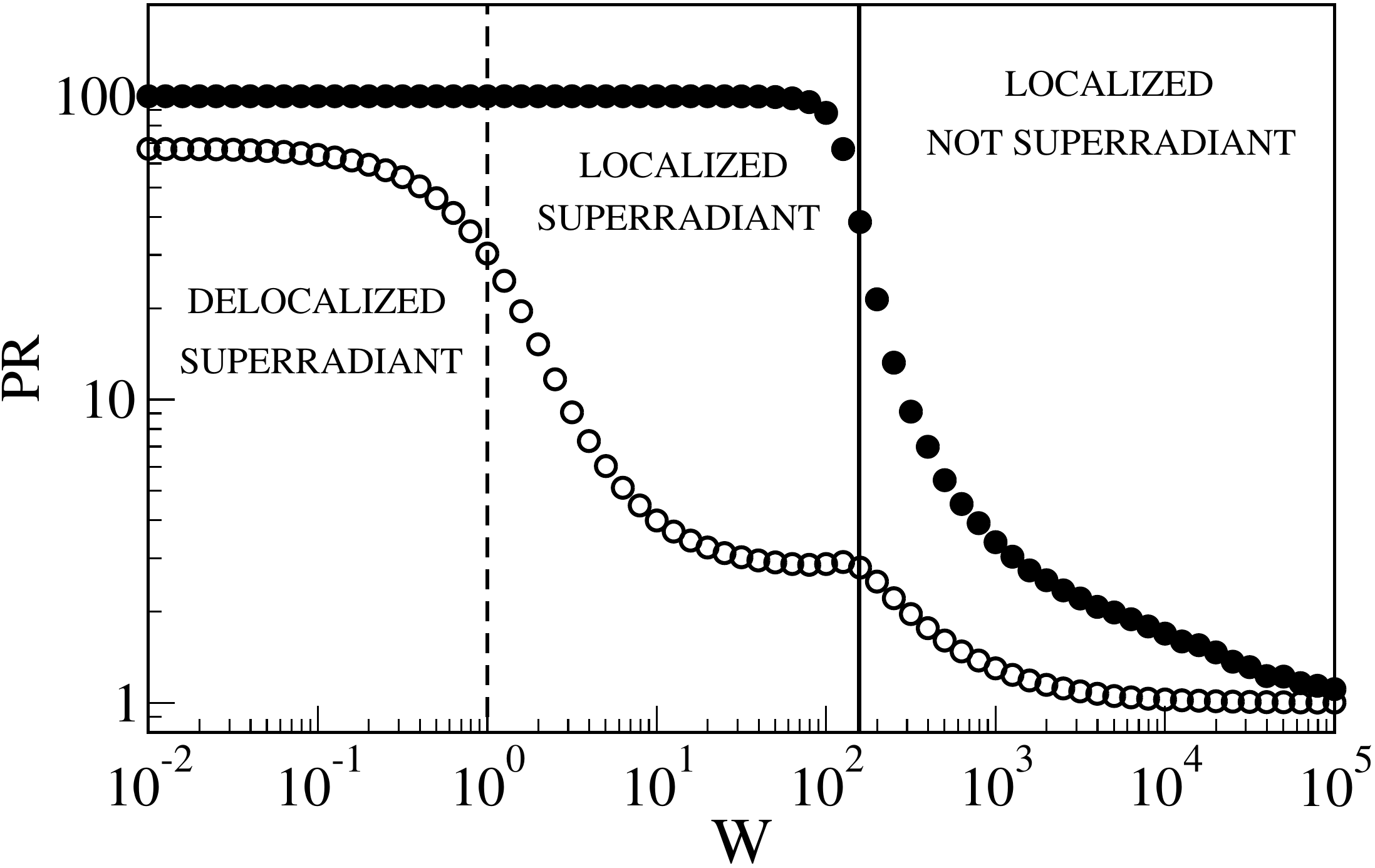}
\caption{The participation ratio is shown as a function of the disorder strength $W$. 
Open  circles stand
 for the subradiant states, while full circles indicate
the superradiant state. Each point is 
obtained by averaging over $100$
 disorder realizations for the superradiant state, while for the
subradiant states, an additional average over all the subradiant states
 is performed. 
The right and left vertical lines indicate the ST and the
delocalization transition, respectively.
Here $N=100$ and $\gamma=\Omega=1$.
}
\label{PRW3}
\end{figure}

We now look more closely at how the subradiant and superradiant states
are affected differently by increasing the disorder strength $W$. 
In Fig.~\ref{PRW3}, we consider the case of $N=100$ and $\gamma=\Omega=1$. 
For small disorder we have   $D \approx 4 \Omega/N$, so that
 $$\kappa=\gamma/D = \gamma N/4\Omega \approx 25 \gg 1\,.$$
This implies that we are in the superradiant regime.
Moreover for sufficiently small disorder we have that the localization length is 
larger than the system size,  $\xi \approx 100 \ \Omega^2/W^2 >N$,
so that both superradiant and subradiant states are delocalized.
 For larger disorder, here  $W>1$, we enter the localized regime, for which
$\xi < N$. In this regime the $PR$ of the subradiant states starts
 to decrease, while the $PR$ of the superradiant state remains 
unchanged ($PR=N$),
 signaling a superradiant state that remains completely delocalized. 
 As we increase disorder further, $\kappa$ decreases according to 
Eq.~(\ref{k}). The ST occurs
at $W \approx \gamma N $, here $W \approx 100$, and above this value 
the superradiance effect disappears. 
Summarizing, we have a critical value of disorder ($W \approx 100$
indicated as a full vertical line in Fig.~\ref{PRW3})  separating
the superradiant regime ($\kappa > 1$), from the 
non-superradiant one ($\kappa<1$).   
Only for $W>100$, i.e., below the ST, do the superradiant states begin
to localize, and, for very large disorder, corresponding 
to very small $\kappa$,
they behave the same as the subradiant states. 

\section{Discussion}
\label{sec-pt}
In this Section we will justify (using perturbation theory) and briefly discuss the interesting results presented
previously: for small $\kappa$ (below the ST) all the states
are affected in a similar way by the opening and disorder,
 while for large $\kappa$ (above the ST),
the superradiant states remain completely delocalized, independent of
the degree of disorder, while the subradiant states
are still sensitive to disorder, and their $PR$
 decreases with increasing disorder.

\subsection{Perturbative approach for $\kappa \ll 1$}

In the limit $\kappa \ll 1$, the eigenstates of $H_{\rm eff}$
at first order in perturbation theory can be written as:
\begin{equation}
| n \rangle  =\frac{1}{\sqrt{C_n}}\left[
 | n^0 \rangle  - 
i \frac{\gamma}{2} \sum_{k^0\neq n^0} \frac{\langle k^0 | Q | n^0 \rangle } {E_{n^0} - E_{k^0}} 
|k^0 \rangle  \right] \,,
\label{pert1}
\end{equation}
where $| n^0 \rangle$ are the eigenstates of the closed system, i.e., of the Anderson model.
Of course, the perturbation expansion makes sense only when each coefficient in the
sum in Eq.~(\ref{pert1}) is much less than one. This cannot be true
in general since the eigenvalues $E_{n^0}$ are random numbers uniformly
distributed in the interval $[-W/2, W/2]$.
Thus perturbation theory cannot be applied {\it tout court},
but only for those states whose energies are not too close.

\begin{figure}[htbp]
\centering
\includegraphics[scale=0.4]{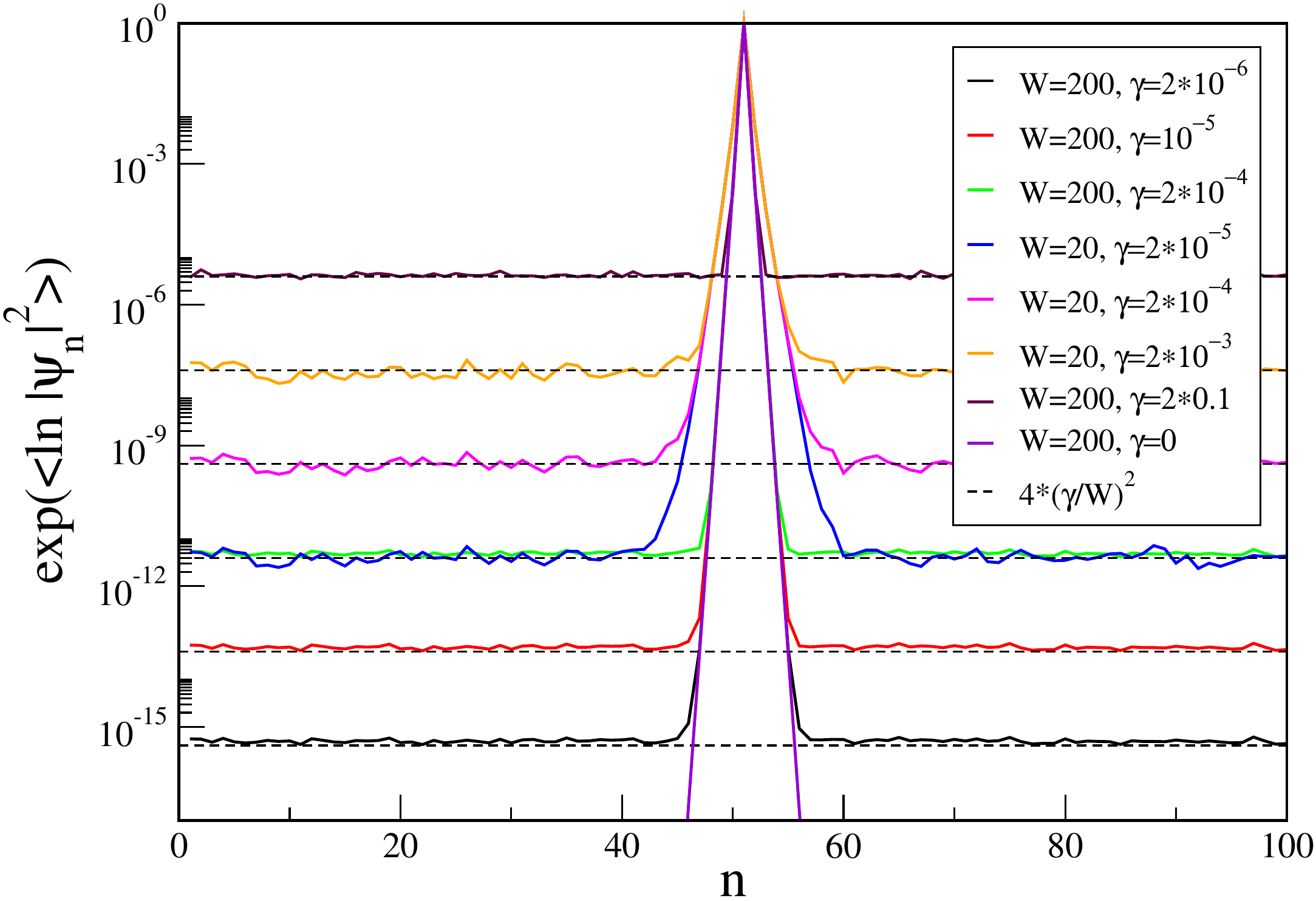}
\caption{The averaged probability distribution of all eigenstates of the non-Hermitian Hamiltonian that are
strongly peaked  in the middle of the chain is shown
for different coupling strength $\gamma$ and disorder strength $W$, as indicated in the caption.
Specifically, we average over all 
eigenstates having a probability $>0.9$ at the site $n=N/2+1$  in order to avoid double-peaked states, and also average over disorder. 
Moreover, to reduce fluctuations, we average the logarithm of
the probability distribution.
In all cases we fix $N=100$ and $\Omega=1$.
Dashed horizontal lines are proportional to $(\gamma/W)^2$ in agreement with
the perturbative approach.
}
\label{PRk}
\end{figure}

This simple observation has deep consequences for the structure
of the eigenstates. Indeed we observe numerically that on the one hand
many single-peaked eigenstates become double- or multiple-peaked
as $\gamma$ increases,
while on the other hand, they all develop a constant plateau
proportional to $(\gamma/W)^2$, see Fig.~\ref{PRk}. 

This last fact can easily be explained using first-order perturbation
theory as given by Eq.~(\ref{pert1}): in the deep localized regime 
$W \gg \Omega$, 
the matrix elements $\langle k^0 | Q | n^0 \rangle$ are of order unity and
the average distance between two random energies is $W/3$, 
so the typical coefficients $\langle k_0|n\rangle$ in Eq.~(\ref{pert1}) 
are $\sim \gamma/W$. 
Furthermore, 
the mean level spacing is $D \approx W/N$, and thus the few largest coefficients 
in Eq.~(\ref{pert1}) are typically $\sim \gamma N/W \sim \kappa$ (using Eq.~(\ref{k})). Thus for weak opening ($\kappa \ll 1$), the {\it typical} eigenstate consists of a single Anderson model eigenstate with a $O(\kappa^2)$ admixture of other states, and therefore the {\it typical} $PR$ for small $\kappa$ differs only by $O(\kappa^2)$ from the $PR$ of the Anderson model.

\begin{figure}[htbp]
\centering
\includegraphics[scale=0.4]{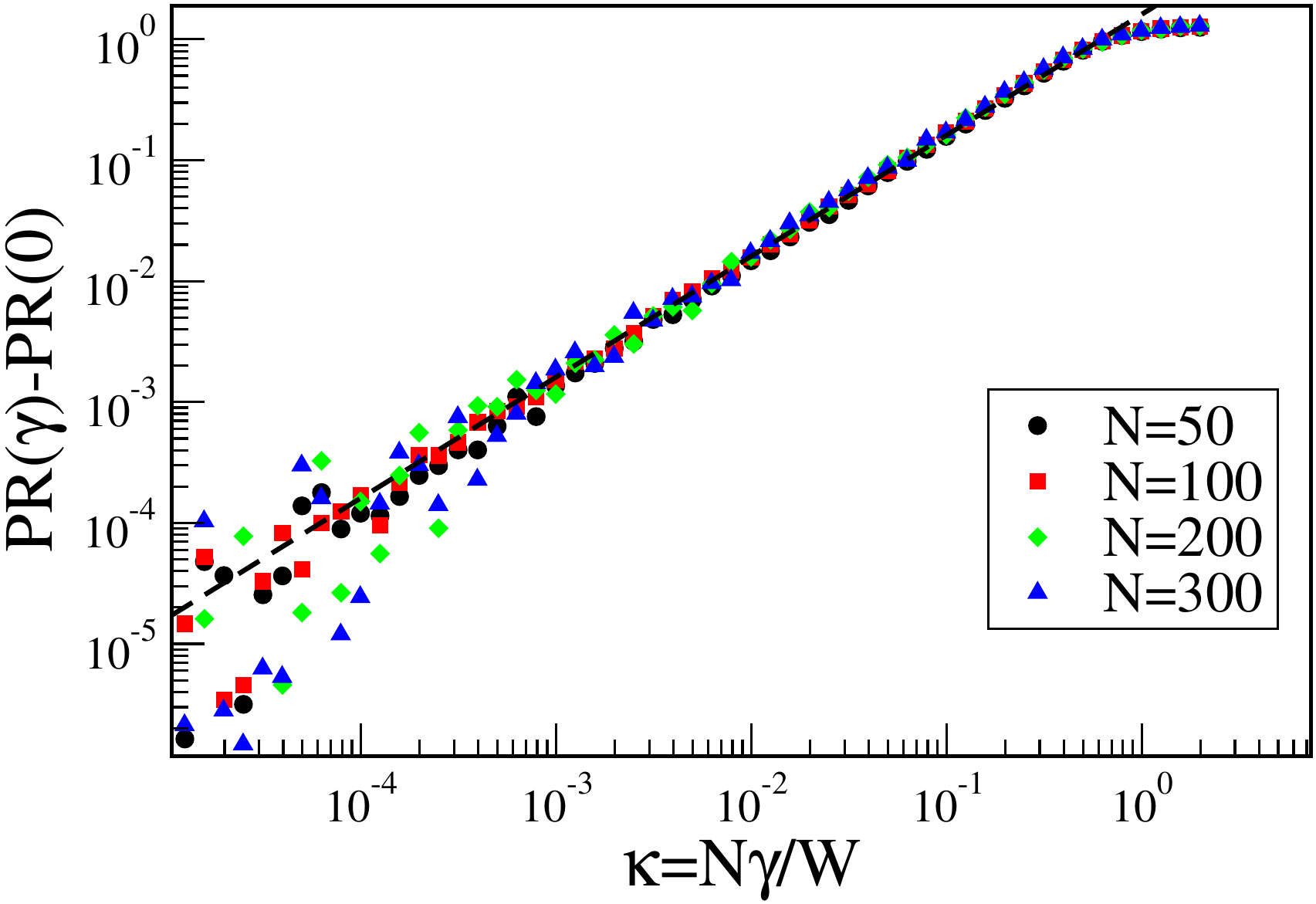}
\caption{The average increase in the participation ratio, compared with the closed
system, is calculated as a function of $\kappa$, for fixed disorder $W=20$ and different system
sizes $N$ as indicated in the legend. In each case the average is performed over 50000 different eigenstates.
The line is $PR(\gamma)-PR(0)=2\kappa$.}
\label{albi}
\end{figure}

As already remarked previously, the perturbative approach
 cannot always work, because for arbitrarily small $\kappa$ 
there is a small but  
finite probability that two energy states are too close 
together. This clustering behavior has important consequences for the localization properties. 
Specifically, since the nearest-neighbor level spacing 
distribution of uniform random numbers $E_{n^0}$ is Poissonian: 
$P(s)=(1/D) \ e^{-s/D}$, where $s$ is the energy difference between 
 nearest-neighbor  levels and $D=W/N$ is the mean level spacing,
we can evaluate the probability to have two levels closer than 
$\gamma/2$ as $1-e^{-\gamma/2D} \approx \kappa/2$ for small $\kappa$. 
This means that there are $\kappa N$ states out of $N$, for which perturbation theory cannot be
applied. When this happens, the Anderson states mix strongly and the $PR$ increases by an $O(1)$ factor. Thus, even though this behavior is rare, it makes an $O(\kappa)$ contribution to the {\it average} $PR$ of the weakly open system, which exceeds the $O(\kappa^2)$ contribution from the typical states.
Indeed the average $PR$ can be evaluated as follow: 
$$
PR= \frac{N\kappa PR_2+ (1-\kappa)N PR_1 }{N}= PR_1 + \kappa (PR_2-PR_1)
$$
where $PR_1$ and $PR_2$ refer to the $PR$ of the states for which
perturbation theory can and cannot be applied.
Since $PR_1 \simeq PR(\gamma=0) + O(\kappa^2)$, and $PR_2 \simeq O(1)$, we have
that  $PR(\gamma) - PR(\gamma=0) \simeq \kappa$.  
The numerical results in Fig.~\ref{albi} confirm that the effect of the opening on the $PR$ grows as $\kappa$, instead of the $\kappa^2$ growth predicted by first-order
perturbation theory.
Here we present the average (over disorder) of $PR(\gamma)-PR(\gamma=0)$,
as a function of 
$\kappa=N\gamma/W$ for fixed disorder strength and different values of the system size.
In any case this is quite a delicate point and we postpone its full analysis
to a future work.

\subsection{Perturbative approach for $\kappa \gg 1$}

In the limit $\kappa \gg 1$ we consider two cases. First, we consider the situation
where the nearest neighbor tunneling 
coupling is $\Omega = 0 $, in which case we can follow the approach explained in Ref.~\cite{sokolov}. 
This approach will be very useful also for the case
$\Omega\neq 0$, which we treat below. 

\subsubsection{$\Omega=0$ {\rm and} $\kappa \gg 1$}
\label{subsub1}
If $\Omega=0$ the Anderson Hamiltonian is diagonal in the site basis $|j\rangle$ with
 eigenvalues $E_j$ distributed uniformly in the interval $[-W/2,W/2]$.
The eigenstates of the non-Hermitian part $-i\frac\gamma2Q$ of the effective Hamiltonian
are $ | d \rangle  =\frac1{\sqrt{N}}(1,...,1)^T$ (the superradiant state) 
with eigenvalue $-i\frac\gamma2N$, and  $N-1$ degenerate eigenstates $| \mu\rangle$ 
with eigenvalue $0$ (the subradiant states). We will choose $| \mu \rangle $ in a convenient manner later.
Following Ref.~\cite{sokolov} we can rewrite $H_{\rm eff}$ in 
the basis of these eigenstates using the transformation matrix $V$, which has
as its columns the eigenstates of $Q$:
\begin{equation}
\tilde{H}_{\rm eff} = V^T H_0 V -\rm i\frac\gamma2V^T Q V = \left(
\begin{array}{cc}
 -\rm i\frac\gamma2N  & \vec{h}^T  \\
\vec{h}  & \tilde{H} \\ 
\end{array}
\right ) \,.
\label{h2}
\end{equation}
Here $\vec{h}$ is a vector of dimension $N-1$ with components
\beq
h_\mu =\frac1{\sqrt{N}}\sum_{j=1}^N E_j  \langle j|\mu \rangle\,,
\eeq
while the matrix elements of the $(N-1)\times(N-1)$ submatrix $\tilde{H}$ are
\beq
\tilde{H}_{\mu\nu}=\sum_{j=1}^N E_j \langle \mu|j \rangle \langle j|\nu \rangle.
\eeq
Now, we can diagonalize $\tilde{H}$,
\beq
\tilde{H}_{\mu\nu}=\sum_{j=1}^N E_j \langle \mu|j \rangle
\langle j|\nu\rangle =\langle \mu|H_0|\nu\rangle = \tilde{\epsilon}_\mu \langle\mu|\nu\rangle.
\eeq
Following Ref.~\cite{sokolov} we obtain 
\beq
|\mu\rangle =h_\mu \frac1{\tilde{\epsilon}_\mu -H_0} |d\rangle= \frac{h_\mu}{\sqrt{N}}\sum_{j=1}^N \frac1{{\tilde{\epsilon}_\mu-E_j}} |j \rangle\,,
\label{sub}
\eeq
where the normalization coefficients $h_\mu$ are given by
\beq
h_\mu=\left(\langle d | \frac{1}{(\tilde{\epsilon}_\mu-H_0)^2} |d\rangle \right)^{-1/2} \,.
\eeq
In the limit $\kappa \gg 1$, the eigenstates $|\mu\rangle$ of the non-Hermitian part of 
$H_{\rm eff}$ are also eigenstates of $H_{\rm eff}$.
Since $\langle d|\mu \rangle =0$ we have,
\beq
\sum_{j=1}^N\frac{1}{\tilde{\epsilon}_\mu-E_j}=0.
\eeq
Therefore  each eigenvalue of $\tilde{H}$ lies between two neighboring
 levels $E_n$, so that the values $\tilde{\epsilon}_{\mu}$ are also confined
in the interval  
$[-W/2, W/2]$. 

Let us now estimate the magnitude of the
mixing matrix elements $h_\mu$. To do this we compute
\beq
\vec{h}\cdot\vec{h} = \frac1N\sum_{\mu=1}^{N-1} \sum_{i=1}^N \sum_{j=1}^N \ E_i E_j
\langle \mu|i \rangle \langle j|\mu\rangle \,,
\eeq
and using the completeness relation $\sum_{\mu=1}^{N-1} \langle j|\mu \rangle 
\langle \mu|i \rangle 
=\langle j|i \rangle-1/N$ we have
\beq
\vec{h}\cdot\vec{h} =  \langle E^2\rangle -\langle {E}^2\rangle = \Delta{E}^2 \,.
\eeq
This leads to
\beq
|h_\mu|\sim \frac{\Delta E}{\sqrt{N-1}} = \frac{W}{\sqrt{12(N-1)}}\,.
\eeq

Each eigenstate $| \mu \rangle $ in Eq.~(\ref{sub}) is a  superposition of all the site states $ |j \rangle $
with amplitudes $\frac{h_\mu}{\sqrt{N}(\tilde{\epsilon}_\mu-E_j)} \sim \frac{W}{N(\tilde{\epsilon}_\mu-E_j)}$ that depend
only on the energies $E_j$ and not on the site positions $j$.
Nevertheless, each state $|\mu\rangle$ is 
quite localized, since the amplitudes are of order unity for the $O(1)$ number of sites whose energy is
within a few mean level spacings of $\tilde{\epsilon}$ (i.e., when $|\tilde{\epsilon}_\mu-E_j |\sim D =W/N$), and small otherwise.
This small value of the $PR$ for the subradiant states should be compared 
with $PR=N$ of the superradiant states.

The values obtained above for the subradiant and the superradiant states
 correspond to zeroth-order perturbation theory. 
On the other hand
first-order perturbation theory gives:
\begin{eqnarray}
&&| SR \rangle = \frac{1}{\sqrt{C}}\left[ |d \rangle
 + \frac{W}{\sqrt{12(N-1)}} \sum_{\mu=1}^{N-1}\frac{r_\mu}{-
i\frac\gamma2N -\tilde{\epsilon}_\mu} | \mu \rangle\right]\cr  
&&\cr
&&=\frac{1}{\sqrt{C}}\left[ |d \rangle + 
  \ \frac1{\kappa \  \sqrt{3(N-1)}} \sum_{\mu=1}^{N-1}\frac{r_\mu}{i-2\tilde{\epsilon}_\mu/\gamma N} 
| \mu \rangle\right]\cr
&&\cr
&&| {SU\!B}_{\mu}\rangle= \frac{1}{\sqrt{C'_\mu}}\left[| \mu \rangle + \frac{W}{\sqrt{12(N-1)}} 
\frac{r_\mu}{\tilde{\epsilon}_\mu+i\frac\gamma2 N} |d \rangle\right]\cr
&&\cr
&&=\frac{1}{\sqrt{C'_\mu}}\left[| \mu \rangle -
 \frac1{\kappa \sqrt{3(N-1)}}\frac{r_\mu}{i-2\tilde{\epsilon}_\mu/ \gamma N} | d \rangle
\right]\,,
\label{firstorder}
\end{eqnarray}
where $r_\mu$ are random coefficients with $\langle r_\mu^2\rangle=1$. 
We see that  the exact superradiant state $| SR \rangle$ is a combination of the 
unperturbed superradiant state $| d \rangle$
 and a small admixture of the unperturbed subradiant states $| \mu \rangle$, and the mixing 
probability decreases as $1/ \kappa^2$ for large $\kappa$. Similarly, the
admixture of the unperturbed superradiant state $| d \rangle$ in
each exact subradiant states $| {SU\!B}_{\mu}\rangle$ decreases as  $1/ (\kappa^2 N)$.
This shows that $PR \approx N$ for the superradiant state and $PR \sim 1$ for the subradiant
states when $\kappa \gg 1$.

\subsubsection{$\Omega \ne 0$ {\rm and} $\kappa \gg 1$}

As a first step we write the Anderson Hamiltonian $H_0$
in terms  of its eigenstates $|n \rangle$.
Obviously the form of $| n \rangle$ will depend on the degree of disorder $W$.
In the following we limit our considerations to the large disorder regime,
so that in the basis of the eigenstates of  $H_0$, 
the matrix elements of $Q$ remain of order one, $Q_{nm} \sim 1$, 
and we can use the results of Sec.~\ref{subsub1}, with the
site states and energies $|j\rangle$ and $E_j$ replaced by the
Anderson eigenstates and eigenenergies $|n\rangle$ and $E_n$.

In Fig.~\ref{PRW3} we see that for $\kappa>1$ (corresponding to
$W<100$), the superradiant state remains unaffected by the increase 
of disorder, while the subradiant states become more localized 
as the disorder strength is increased.  
The results of the previous section can be used to understand this strongly 
asymmetric behavior of the $PR$ between the subradiant states and the 
superradiant state. 
Indeed at zeroth order in perturbation theory we can see that the
 superradiant state $|SR\rangle \approx |d\rangle$ is completely delocalized, $PR= N$, while
 subradiant states $| {SU\!B}_{\mu}\rangle \approx |\mu\rangle$
 become more and more localized as we increase disorder.
Specifically, the site states $|j\rangle$ in Eq.~(\ref{sub})
are replaced with Anderson eigenstates $|n\rangle$, with localization length  $ \xi \propto 1/W^2$.  
This difference persists in first-order perturbation theory, since the mixing probability between the super- and sub-radiant states
decreases as $1/\kappa^2$ for large $\kappa$, see Eq.~(\ref{firstorder}).

Our perturbative approach justifies the results presented in Fig.~\ref{PRW3},
where we  can see  that 
the subradiant states become increasingly localized 
as we increase disorder.   
At the same time Fig.~\ref{PRW3} shows that
the superradiant state remains completely delocalized even as $W$ 
is increased, until
we reach the value $W\approx142.8$ ($\kappa=0.7$) where we find 
numerically that the ST
takes place. 
The perturbative approach shows
 that superradiant states are much less sensitive to disorder
because their complex energies
are at a distance greater than  $\gamma N/2 = W \kappa/2 $  
from the subradiant states.

\section{Conclusions}
\label{sec-conc}
We have studied a $1$-D Anderson model with all sites
coupled to a common decay channel (coherent dissipation). 
Our main motivation  was to understand the interplay of opening 
and disorder. 
  Increasing the disorder tends to localize the states. Increasing the opening, on the other hand, reduces the degree of localization, and in particular induces a superradiance transition, with the formation of a subradiant subspace and a superradiant state completely delocalized over all the sites.
Our results show that, while for
small opening all the states tend to be similarly affected by the 
disorder, for large opening the superradiant state remains delocalized
even as the disorder increases, while the subradiant states
are much more affected by disorder, becoming ever more
localized as the disorder increases. We have explained these effects qualitatively,
mainly guided by  perturbation theory. Indeed we have shown that 
the superradiant state is not affected by disorder, since its energy
is very distant, in the complex plane,  from the energies of
the subradiant states.


\section*{Acknowledgments}

We acknowledge useful discussions with  I.~Rotter, R.~Kaiser, and V.~G.~Zelevinsky. 
This work has been supported by Regione Lombardia and CILEA Consortium through a LISA Initiative (
Laboratory for Interdisciplinary Advanced Simulation) 2011 grant [link:http://lisa.cilea.it]. 
 Support by the grant D.2.2 2011
(Calcolo ad alte prestazioni)  from Universit\'a  Cattolica
 is also acknowledged.


\end{document}